\documentclass{PoS}

\usepackage{graphicx}

\usepackage{epsfig}

\usepackage{multicol}
\usepackage{rotating}

\title{Interpretation of charged Higgs effects in low energy flavour physics\footnote{MZ-TH/11-06}}

\ShortTitle{Charged Higgs  effects in low-energy  flavour physics}

\author{\speaker{Tobias Hurth}\\
       Institute for Physics, Johannes Gutenberg-University, D-55099 Mainz, Germany\\         
 E-mail: \email{tobias.hurth@cern.ch}}

\abstract{We discuss  two-Higgs-doublet models  in view of the present flavour data, in particular
present indirect bounds and different techniques of flavour protection.}

\FullConference{Third International Workshop on Prospects for Charged Higgs Discovery at Colliders - CHARGED2010,\\ invited plenary talk, 
		September 27-30, 2010\\
		Uppsala,  Sweden}

\begin{document}

\section{Introduction}

The two-Higgs-doublet model (THDM) constitutes  one of the simplest extensions of the present
Standard Model (SM).
Many new-physics scenarios, including supersymmetry, can lead to a low-energy spectrum containing  the SM fields plus at least one additional scalar doublet. 
Also recent developments in string phenomenology indicate that an additional generation of Higgs bosons  is generic within this framework~\cite{ Gupta:2009wn}. 

From the two Higgs doublets, three degrees of freedom are {\it eaten} and become longitudinal 
components of the gauge bosons and five degrees are left: the scalar mass eigenstates, $H, h$, the pseudoscalar $A$, and the charged Higgs bosons $H^\pm$.  These new degrees of freedom 
induce new flavour-changing neutral currents in general. In consequence, an analysis of this class of models in  view of the huge data sets from the recent flavour experiments is desirable.

The first generation of the $B$ factories at KEK (Belle experiment at
the KEKB $e^+ e^-$ collider)~\cite{Belle} and at SLAC (BaBar experiment at
the PEP-II $e^+ e^-$ collider)~\cite{Babar}   have collected huge samples of
$B$ meson decays and thus established the SM picture of CP
violation and other flavour-changing processes in the quark sector.
It is remarkable that all present measurements of $B$ meson decays (including the $B$ physics programme  at the Tevatron (CDF~\cite{TevatronB1} and
D0~\cite{TevatronB2})  have not
observed any unambiguous sign of
new physics (NP)  yet~\cite{Buchalla:2008jp,Antonelli:2009ws}.  

%%%%%%%%
\begin{figure}[htb]
\begin{center}
\includegraphics[width=1.00\textwidth]{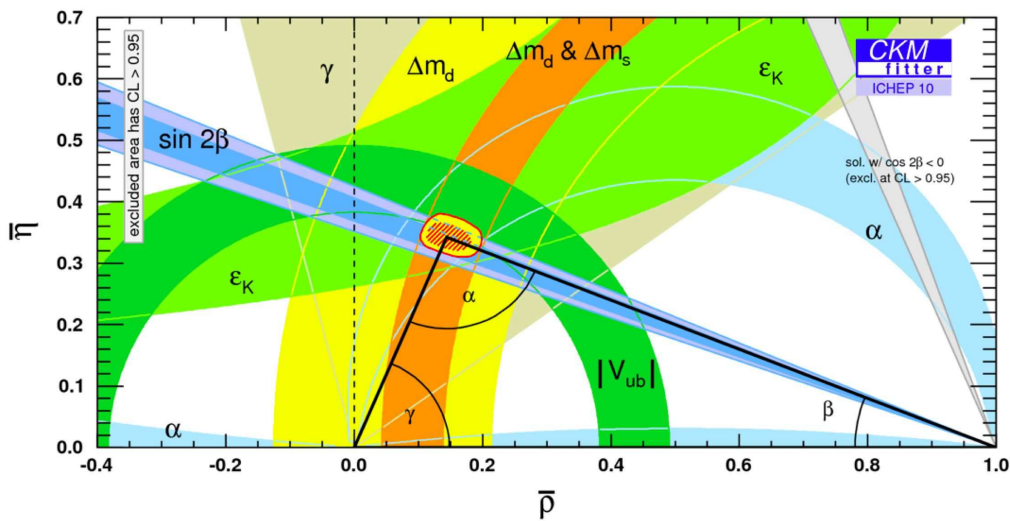}
\end{center}
\caption{CKM unitarity fit~\cite{CKMfitter}.}
\label{fig:CKM1}
\end{figure}
%%%%%%%%%%%%

\begin{figure}[htb]
\hspace{-1cm}\includegraphics[width=.55\textwidth]{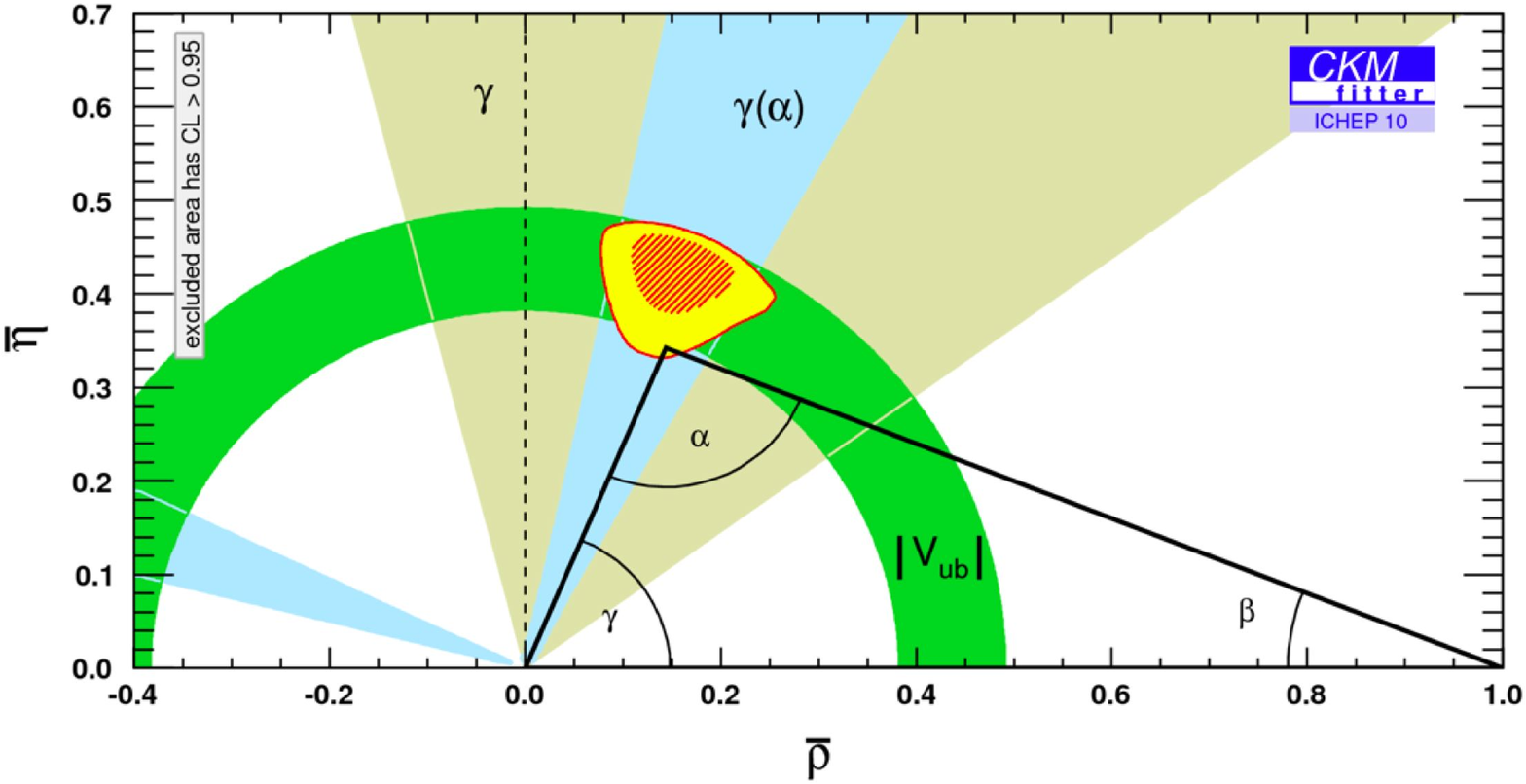}\hspace{-0.4cm}
\includegraphics[width=.55\textwidth]{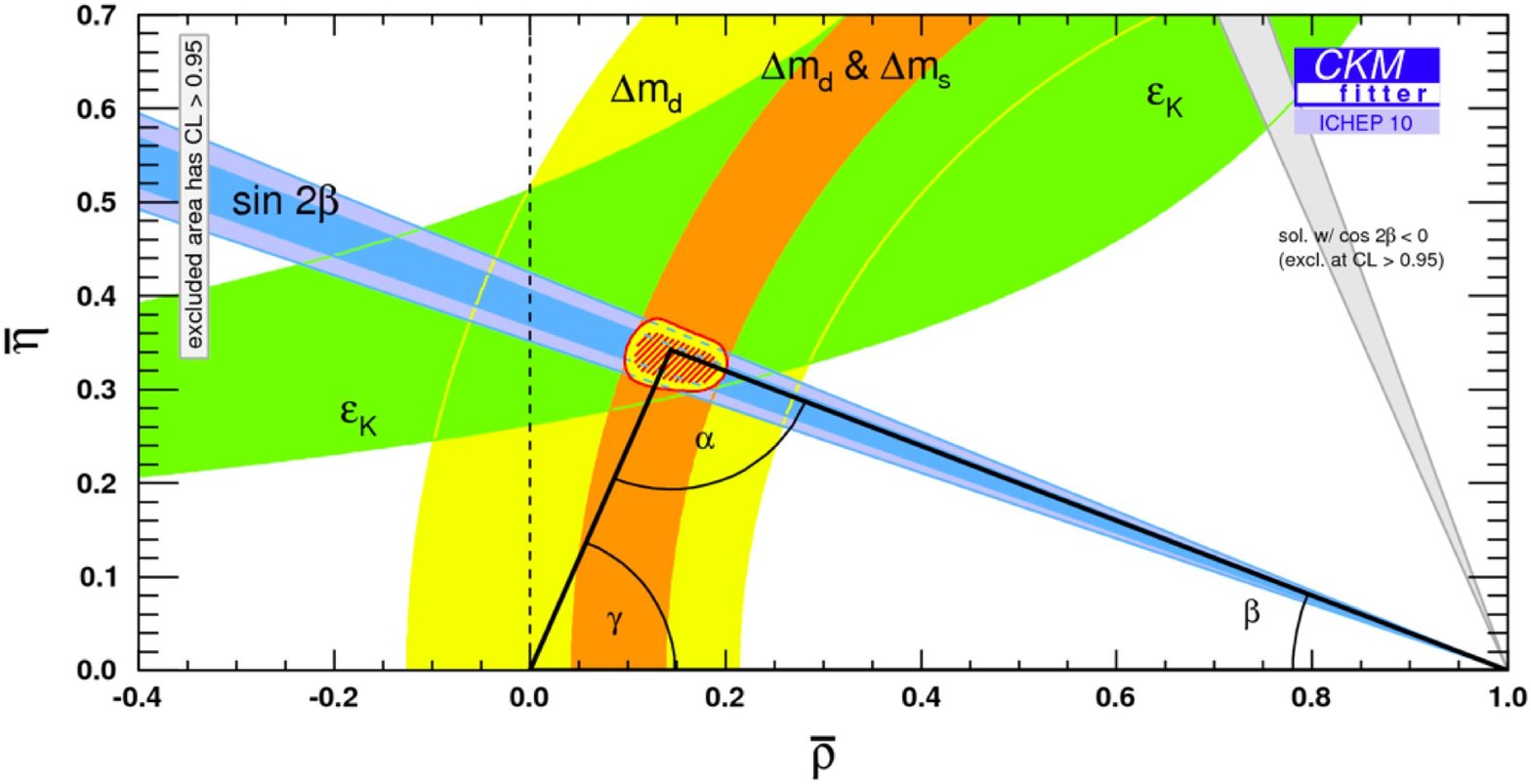}
\caption{Unitarity triangle fixed by tree (left) versus  loop (right) processes~\cite{CKMfitter}.}
\label{fig:CKM23}
\end{figure}

\begin{figure}[htb]
\hspace{-1cm}\includegraphics[width=.55\textwidth]{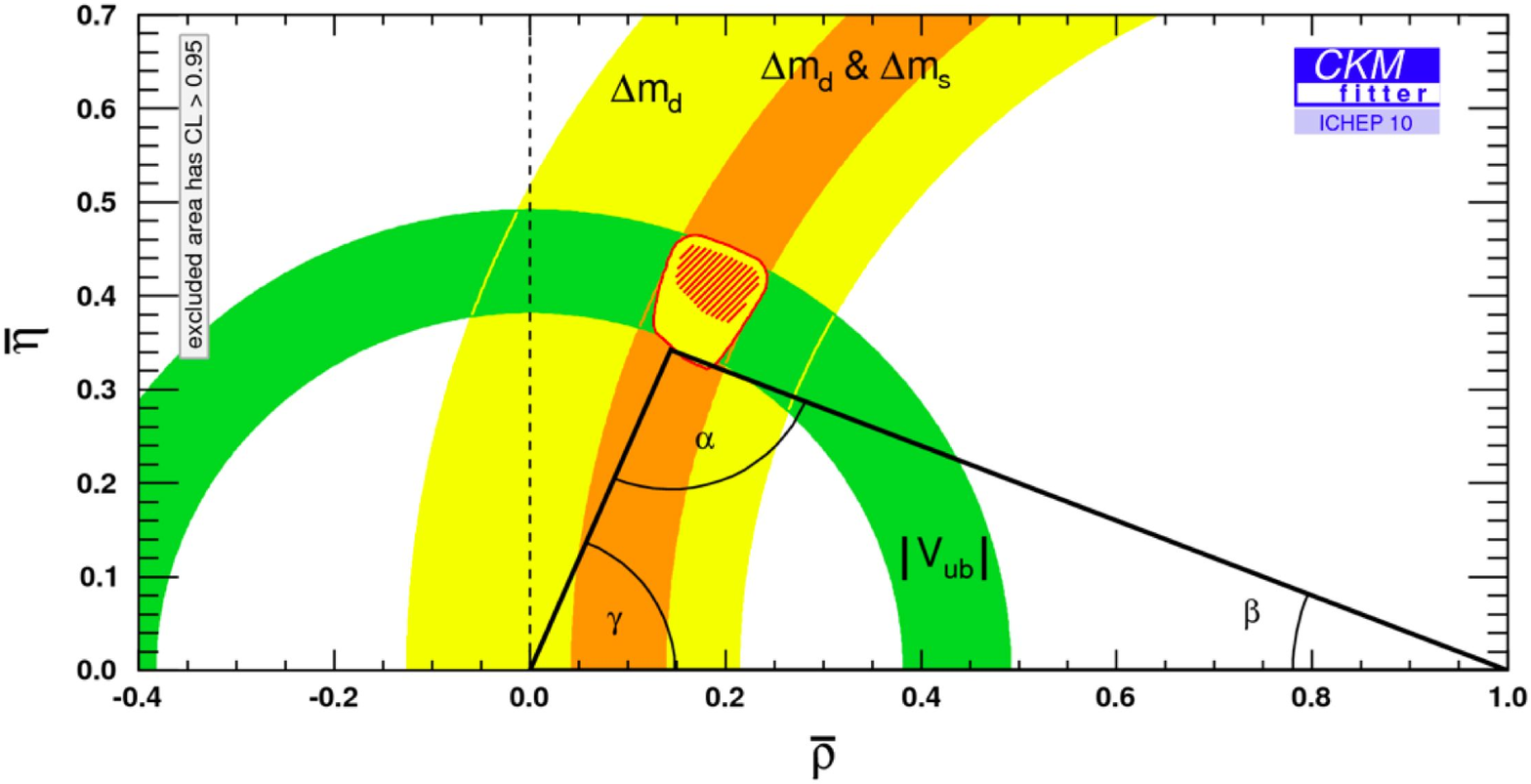}\hspace{-0.4cm}
\includegraphics[width=.55\textwidth]{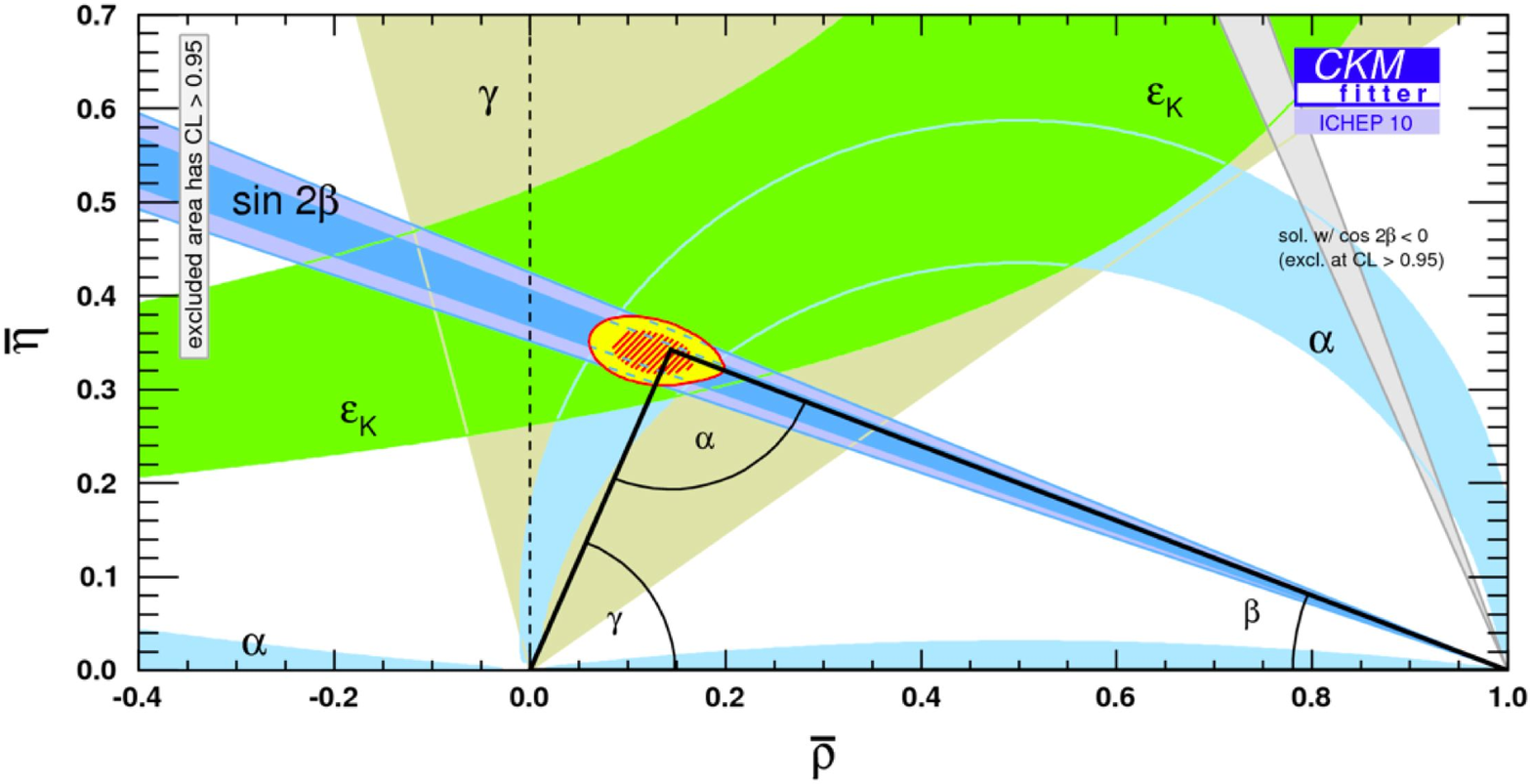}
\caption{Unitarity triangle fixed by CP-conserving (left) versus CP-violating (right) 
processes~\cite{CKMfitter}.}
\label{fig:CKM45}
\end{figure}

This means that
all flavour-violating processes between quarks are governed by a $3
\times 3$ \mbox{unitarity} matrix
referred to as the Cabibbo-Kobayashi-Maskawa (CKM)
matrix~\cite{Kobayashi:1973fv,Cabibbo:1963yz}.  The CKM matrix is fully
described by four real parameters, three rotation angles and one complex
phase. It is this complex phase that represents the only source of CP
violation and that allows a unified description of all the CP-violating
phenomena in the SM.  This can be illustrated by the overconstrained
\mbox{triangle}  in the complex plane which reflects the unitarity of the CKM
matrix, see Fig.\ref{fig:CKM1}.  What is most impressing is the consistency
between tree-level and loop-induced processes. This is remarkable because in 
the latter ones  possible new degrees of freedom might contribute while tree
processes are fully dominated by SM physics, see Fig.~\ref{fig:CKM23}.
One also finds consistency when one separates CP-violating and CP-conserving observables, 
see Fig.~\ref{fig:CKM45}. As we will see in Sect.3, there is much more data not shown 
in the unitarity  fit which confirms the SM predictions like rare decays.  

%%%%%%%%%%%%%%
%%%%%%%%%%%%%%

%%%%%%%%%%%%%%
%%%%%%%%%%%%%%

While this is an impressing success  of the CKM theory~\cite{ Kobayashi:1973fv, Cabibbo:1963yz}
within the SM  which  was honored by  the Nobel  Prize in physics in 2008,
there is still  room for sizeable new effects from new flavour structures (see i.e. Refs.~\cite{Hurth:2003th, Hurth:2009ke}), as FCNC
processes have been tested only up to the $10\%$ level.

The non-existence of large NP effects in the FCNC processes implies the
famous flavour problem, namely why FCNC are suppressed. 
This has to be solved in any viable NP model.
Either the mass scale of the new degrees of freedom is very high or the 
new flavour-violating couplings are small for reasons that remain to be found. 
For example,  assuming  {\it generic\/}  new flavour-violating couplings,  the present data on  
$K$-$\bar K$ mixing implies  a very high NP scale of order $10^3$--$10^4$ TeV 
depending on whether the new  contributions enter at loop- or at tree-level.   
 In contrast, theoretical
considerations on the Higgs sector, which is responsible for the mass generation 
of the fundamental particles in the SM, call for NP at
order $1$ TeV.    As a consequence,
any NP below the 
$1$ TeV scale must have a non-generic flavour structure.  
Moreover, the present measurements of $B$ decays, especially of FCNC
processes, already significantly restrict the parameter
space of NP models.

There has been an intense discussion on how the flavour problem can be solved within the class
of models with two or more Higgs doublets~\cite{Pich:2009sp,Joshipura:2009ej,Botella:2009pq,Tuzon:2010vt,Ferreira:2010xe,Blechman:2010cs,Buras:2010mh}. In a first step it is important to find conditions 
to avoid FCNC at the tree level. However, this might not be sufficient.  It is also important to address 
the question of  the stability of such flavour-protecting conditions under radiative corrections  as was pointed out  most recently~\cite{Buras:2010mh}. 

This  paper is organized as follows. In Section 2 we recall the various  conditions on the THDM which avoid FCNC at the tree level. In Section 3 we discuss present bounds on such models from flavour data
and compare them with information we get from the direct search for NP in high-energy
experiments. In Section 4 we discuss the hypothesis of minimal flavour violation while in the last section we analyze  its  application
to THDM and the stability of the various flavour-protecting conditions.

\section{Tree-level FCNC in the THDM}

In the most general version of the THDM, the fermionic couplings of the neutral scalars are non-diagonal in flavour leading to FCNC at the tree level. In fact, the most general renormalizable
Yukawa interaction reads
\begin{equation}
- L_Y^{\rm general} = \bar Q_L X_{d1} D_R H_1 + \bar Q_L X_{u1} U_R H_1^c 
+ \bar Q_L X_{d2} D_R H_2^c + \bar Q_L X_{u2} U_R H_2 +{\rm h.c.}~,
\label{Lagrangian1}
\end{equation}
with general  $3 \time 3$ flavour matrices and $H_i^c= i \tau_2  H_i^*$. 
The corresponding mass
matrices are given by  
\begin{equation}
M_d = \frac{1}{\sqrt{2}} \left( v_1 X_{d1} + v_2 X_{d2} \right)~, \qquad 
M_u = \frac{1}{\sqrt{2}} \left( v_1 X_{u1} + v_2 X_{u2} \right)~.
\label{Massmatrices}
\end{equation}
where  $v_1$ and $v_2$ are the two vacuum expectation values, $\prec H_i^\dagger H_i \succ = v_i$. 
As already recognized in Refs.~\cite{Glashow:1976nt,Paschos:1976ay}, 
the mass matrices and the physical
couplings to the physical scalar Higgs bosons cannot be diagonalized simultaneously
in general. 
In fact,  after rotating the basis of the Higgs fields, $(H_1, H_2)$ by  the angle $\beta=\arctan(v_2/v_1)$
to the new basis $(S_1,S_2)$, one finds $\prec S_1^\dagger S_1 \succ = v = \sqrt{v_1^2 + v_2^2}$ and  
$\prec S_2^\dagger S_2  \succ = 0$. The Lagrangian separates into the mass terms and the interaction terms to the physical Higgs fields:  
\begin{equation}
-  L_Y^{\rm general} = \bar Q_L ( \sqrt{2}/v\, \,  M_{d}  S_1  + Z_{d}  S_2  )  D_R 
+ \bar Q_L ( \sqrt{2}/v\,\,  M_{u}  S_1^c + Z_{u} S_2^c )  U_R  +  {\rm h.c.}.
\label{Lagrangian2}
\end{equation}
with 
\begin{equation}
Z_d = \cos \beta X_{d2} - \sin \beta X_{d1}, \qquad    Z_u = \cos \beta X_{u2} - \sin \beta X_{u1}.
\label{Couplings}
\end{equation}
It is obvious that for general flavour matrices $X_i$, the matrices $M_u$ and $Z_u$ 
(analogously $M_d$ and $Z_d$) cannot be simultaneously diagonalized. This directly 
generates tree-level  FCNC.

This feature can be cured by the assumption that only one Higgs field can couple to a given
quark species, corresponding to the condition  $X_{u1}= X_{d2} =0$~\cite{Glashow:1976nt,Paschos:1976ay}.
Another possibility to induce the same effect is the setting  $X_{u2}= X_{d2} =0$, which means that 
all fermions couple only to one of the Higgs fields. The corresponding
models are called THDM Type-II and Type-I repectively.  These conditions directly imply 
$M_{u,d} \sim Z_{u,d}$ which prevents the models from FCNC at the tree-level.

The conditions can be implemented by flavourblind discrete  symmetries.  
For the Type-I model the $Z_2$ symmetry is just the exchange symmetry $H_2 \leftrightarrow -H_2$,
with all other fields unchanged,  while  for the Type-II model the $Z_2$  symmetry transformation is 
$H_1 \leftrightarrow - H_1,\, D_R \leftrightarrow  - D_R$.  The latter symmetry 
can be regarded as a 
subgroup of the well-known Peccei-Quinn $U(1)$ symmetry~\cite{Peccei:1977ur}. 
Indeed, the $U(1)_{\rm PQ}$ can be defined within this context  by attributing $D_R$ and $H_1$ 
charges $+1$ and $-1$, respectively and no charge to all other fields.  Then it is clear  that
also the flavourblind Peccei-Quinn $U(1)$ symmetry implies the Type-II model. 

There is yet another more general flavour-protecting condition on the tree level~\cite{Pich:2009sp,Tuzon:2010vt}.  
Moreover, the caveat of all these tree-level implementations 
is their instability under quantum corrections which might not assure sufficient flavour 
protection~\cite{Buras:2010mh}. Both issues will be discussed in the last section. 

\section{Parameter bounds on the THDM}

Rare $B$ and kaon decays (for reviews see \cite{Hurth:2010tk,Hurth:2007xa,Hurth:2003vb}) representing   loop-induced or helicity-suppressed processes
are highly sensitive probes for new degrees of freedom beyond the SM
establishing an alternative way to search for NP. 
The day the existence of new degrees of
freedom is established by the direct search via the Large Hadron Collider (LHC), the
 present stringent flavour bounds will translate in first-rate
information on the NP model at hand.

At present there are two key observables constraining the charged Higgs sector in THDM, the inclusive
$\bar B \rightarrow X_s \gamma$ decay and the leptonic decay $B \rightarrow \tau \nu$.

%%%%%%%%%%
\begin{figure}[htb]
\begin{center}
\includegraphics[width=.50\textwidth]{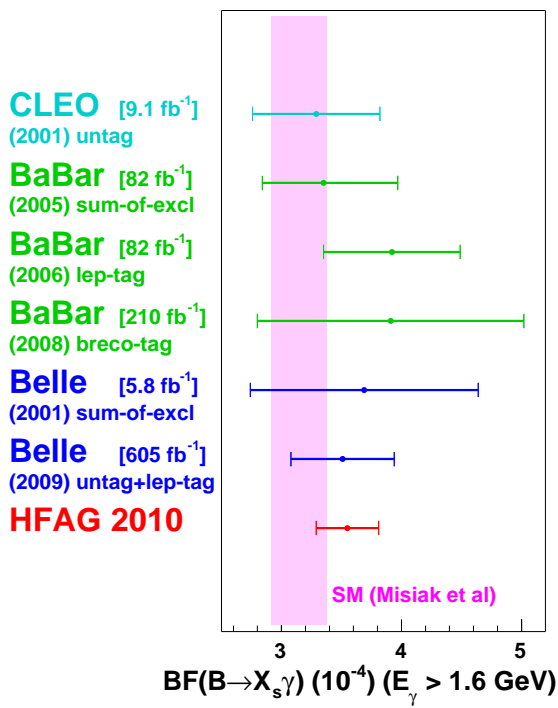}
\caption{Experimental measurement versus NNLL theory prediction of    ${\cal B}(\bar B \rightarrow X_s \gamma)$ ~\cite{Nakao}.}
\end{center}
\label{fig:ttt1}
\end{figure}
%%%%%%%%%%%%%%

Among the rare decay modes, the inclusive decay $\bar B \rightarrow X_s \gamma$  is the 
most important one, because it is theoretically 
well-understood and at the same time it  has been
 measured extensively at the $B$ factories.
While non-perturbative  corrections to this decay mode are subleading and recently estimated to be well below $10\%$~\cite{Benzke:2010js}, 
perturbative QCD corrections are the most important corrections.
Within a global effort,  a  perturbative QCD calculation to the next-to-next-to-leading-logarithmic 
order level (NNLL) has quite recently been
performed and has led to the first NNLL prediction of the $\bar B \to X_s  \gamma$ branching 
fraction~\cite{Misiak:2006zs} with a photon cut at $E_\gamma = 1.6 {\rm GeV}$ (including the error due to nonperturbative corrections):
\begin{equation}\label{final1}
{\cal B}(\bar B \to X_s \gamma)_{\rm NNLL} =  (3.15 \pm 0.23) \times 10^{-4}.
\end{equation}
The combined experimental data leads to 
(Heavy Flavor  Averaging Group (HFAG)~\cite{hfag}) 
\begin{equation}
 {\cal B}(\bar B \rightarrow X_s \gamma) = (3.55
  \pm 0.24 \pm 0.09) \times 10^{-4}, 
\end{equation}
where the first error is combined statistical and systematic, and the second
is due to the extrapolation in the photon energy.
%~\footnote{The HFAG number in 2009 
%has been slightly different: ${\cal B}(\bar B \rightarrow X_s \gamma) = (3.52
 % \pm 0.23 \pm 0.09) \times 10^{-4}$.}.  
Thus, the SM prediction and the experimental average are consistent at the $1.2 \sigma$ level,
see Fig.~4.\,   %\ref{fig:ttt1}.  
This is one important  example that the CKM theory is not only confirmed by 
the data entering into the CKM unitarity fit, but also by many additional flavour mixing phenomena.

%%%%%%%%%%
\begin{figure}[htb]
\begin{center}
\includegraphics[width=.55\textwidth]{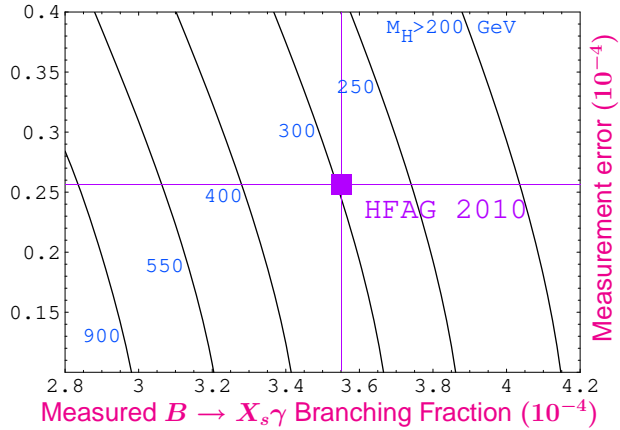}
\caption{Bound on charged Higgs mass depending on the measured branching ratio and on the experimental error~\cite{Nakao}.}
\end{center}
\label{fig:CHMass}
\end{figure}
%%%%%%%%%%%%%%

This specific result in the case of ${\cal B}(\bar B \rightarrow X_s \gamma)$
implies very stringent constraints on NP physics models.  Such bounds are of course
model-dependent, but  in general much stronger than the ones derived from other
measurements. In any case, the indirect flavour information will be most valuable when the
general nature and the mass scale of the NP will be identified in the direct search. 
 
For example one finds a  bound on the inverse
compactification radius of the minimal universal extra dimension model
(mACD) ($1/R > 600 {\rm GeV}$ at $95\%$ CL)~\cite{Haisch:2007vb}.
For the  the two-Higgs doublet model (Type-II), one finds an upper bound 
for the charged Higgs mass, $M_{H^+} > 295{\rm GeV}$
at $95\%$ CL~\cite{Misiak:2006zs}, see Fig.~5.%\ref{fig:CHMass}.

%%%%%%%%%%
\begin{figure}[htb]
\begin{center}
\includegraphics[width=.35\textwidth]{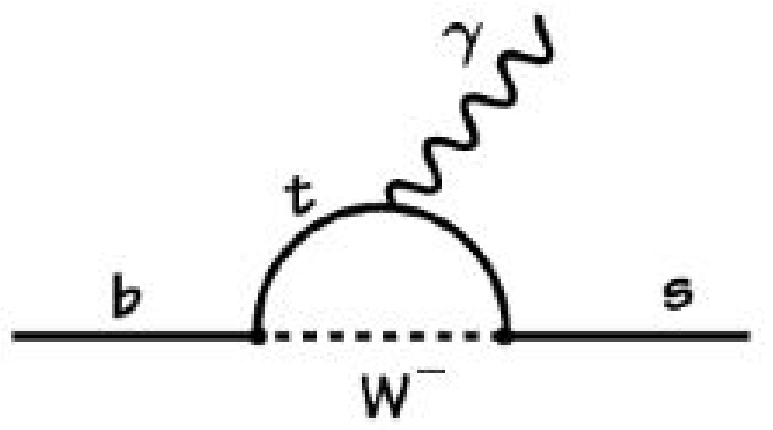}\includegraphics[width=.33\textwidth]{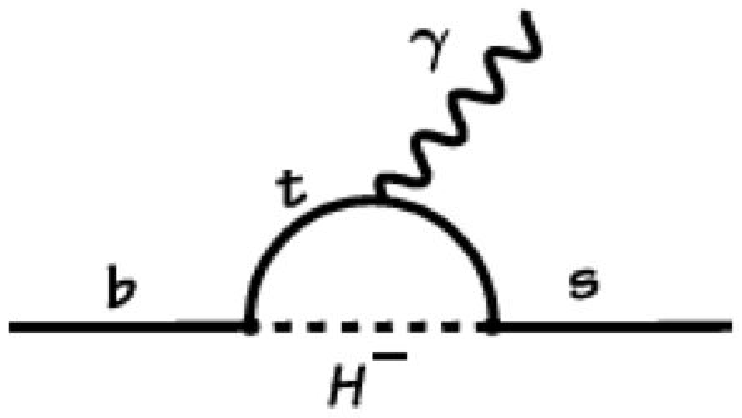}
\caption{SM (left) and charged Higgs contribution (right) to ${\cal B}(\bar B \rightarrow X_s \gamma)$. }
\end{center}
\label{fig:Feynman12}
\end{figure}
%%%%%%%%%%%%%%

%%%%%%%%%%
\begin{figure}[htb]
\begin{center}
\includegraphics[width=.40\textwidth]{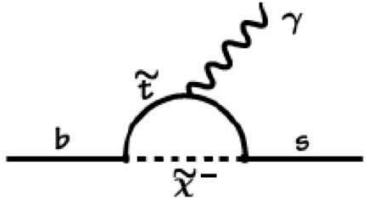}
\caption{Chargino  contribution to  ${\cal B}(\bar B \rightarrow X_s \gamma)$.}
\end{center}
\label{fig:Feynman3}
\end{figure}
%%%%%%%%%%%%%%

In the SM the $b \rightarrow s \gamma$ transition is a loop-induced decay via an exchange of a 
$W$ boson and a top quark, see left diagram of  Fig.~6.\, 
In the THDM-II  there is an additional contribution due to the charged Higgs  (see right diagram 
of Fig.~6)\, %\ref{fig:Feynman12})
which always adds to the SM one which implies the stringent lower bound on the charged Higgs mass
~\cite{Ciuchini:1997xe,Borzumati:1998tg}.
However, embedding the THDM-II into a supersymmtric model, one finds the  chargino contribution
due to quark mixing  (see Fig.~7)\,%\ref{fig:Feynman3}) 
which in principle can destructively interfere 
with the charged Higgs contribution. As a consequence there is no significant bound 
on the charged Higgs mass within the minimal supersymmetric SM (MSSM), see i.e. Refs.~\cite{Bertolini:1990if, Degrassi:2000qf,Carena:2000uj,Degrassi:2006eh}.

Moreover,  the MSSM allows for generic new sources of flavour violation beyond the
CKM structure in the SM. Next to the usual quark mixing also the squark mixing induces flavour mixing
due to a possible misalignment of quarks and squarks in flavour space, see i.e. Refs.~\cite{Borzumati:1999qt, Besmer:2001cj, Ciuchini:2002uv}

The other key observable is the leptonic decay $B \rightarrow \tau \nu$. It is induced in the SM at the tree
level, see left diagram of Fig.~8.\,%\ref{fig:Feynman45}. 
The charged Higgs contribution  modifies
the SM branching ratio as follows:
%%%%
\begin{equation}
{\cal B}_{\rm THDM-II}(B \rightarrow \tau\nu) = {\cal B}_{\rm SM}(B \rightarrow \tau\nu)  \times (1 -  \frac{M_B^2}{M^2_{H^+}}\, \tan^2  \beta) 
\label{btaunu}
\end{equation}
%%%
Thus, the measurement of ${\cal B}(B \rightarrow \tau \nu)$  also implies stringent bounds on the charged Higgs mass depending of the value of $\tan \beta$. 
Complementary, one gets similar information via the measurement of  
${ \cal B} (B \rightarrow D \tau \nu)$, see right diagram of Fig.~8:%\ref{fig:Feynman5}:
%%%%
\begin{equation}
{\cal B}_{\rm THDM-II}(B \rightarrow D \tau\nu)= G_F^2 \tau_B |V_{cb}|^2 \,  f [F_V,F_S, 1 -  
\frac{M_B^2}{M^2_{H^+}}\, \tan^2  \beta]
\label{bDtaunu}
\end{equation}
%%%
The hadronic formfactors  $F_V$ and $F_S$   can be studied via the decay $B \rightarrow D \ell \nu$.

%%%%%%%%%%%
\begin{figure}[htb]
\begin{center}
\includegraphics[width=.40\textwidth]{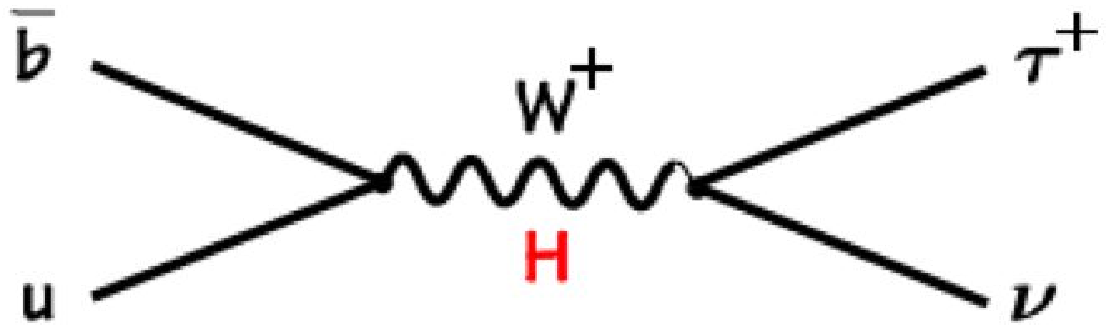}\includegraphics[width=.40\textwidth]{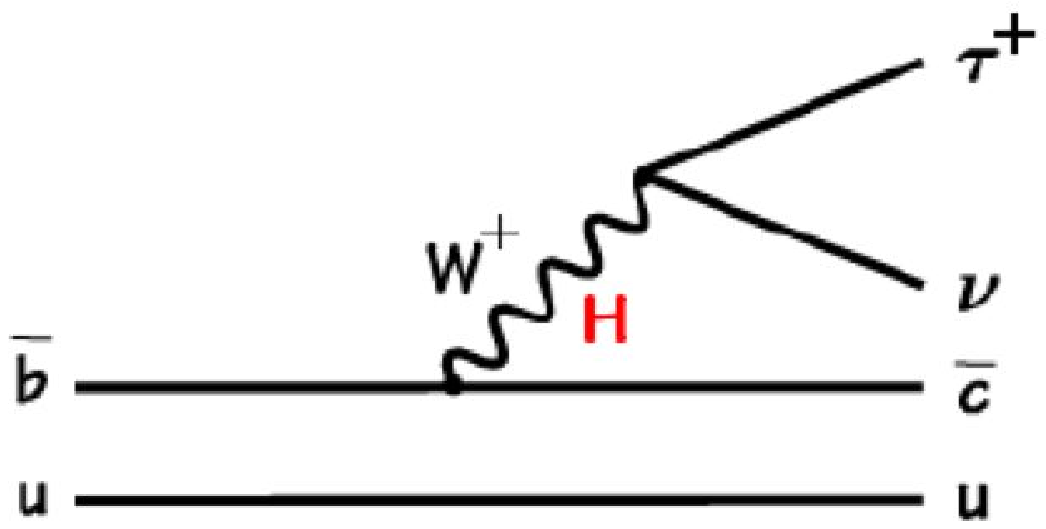}
\end{center}
\caption{Tree contributions to ${\cal B}(B \rightarrow \tau \nu)$ (left) and to ${\cal B}(B \rightarrow D \tau\nu)$ (right).}
\label{fig:feynman45}
\end{figure}
%%%%%%%%%%

\begin{figure}[htb]
\begin{center}
\includegraphics[width=.47\textwidth]{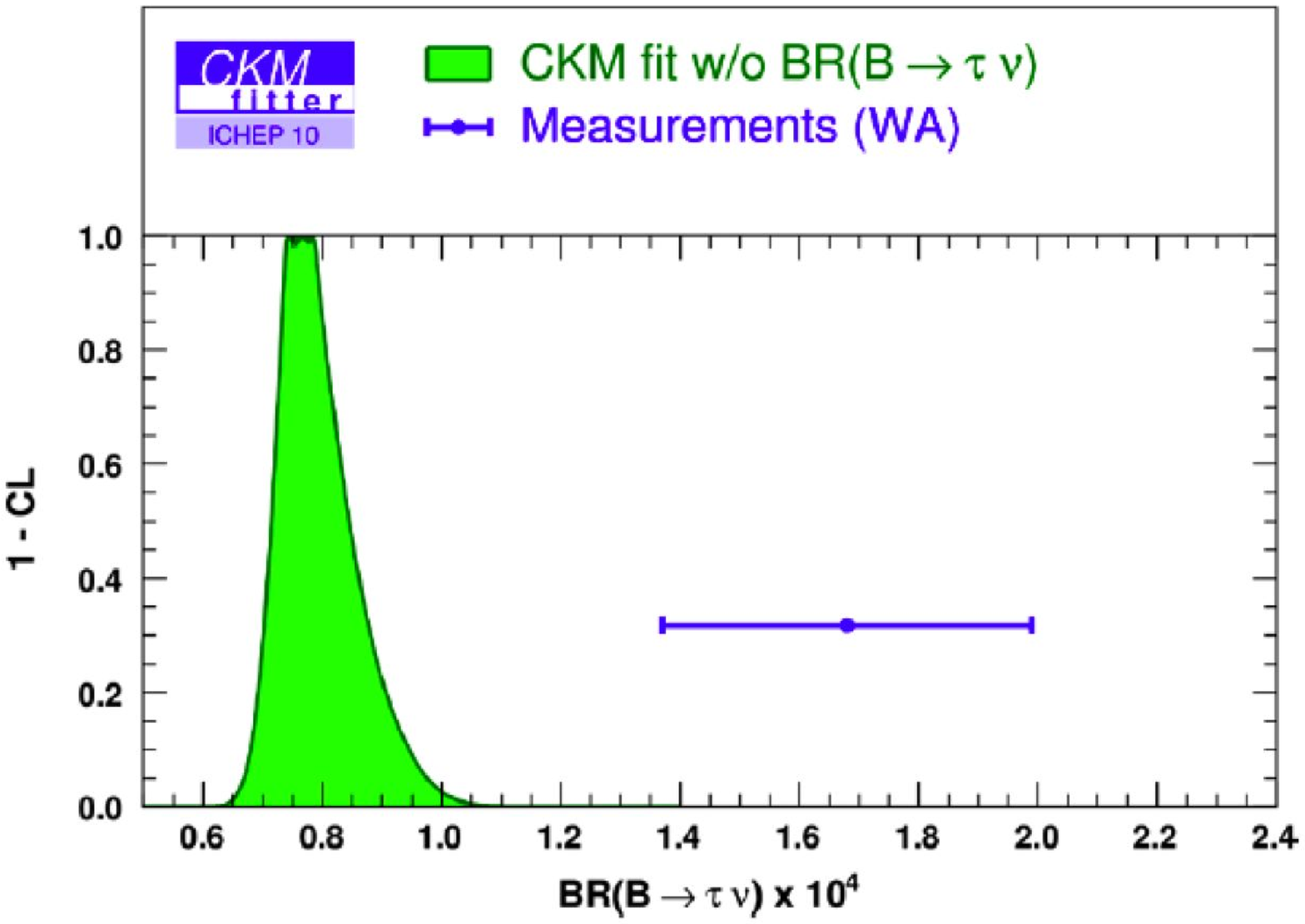}\includegraphics[width=.47\textwidth]{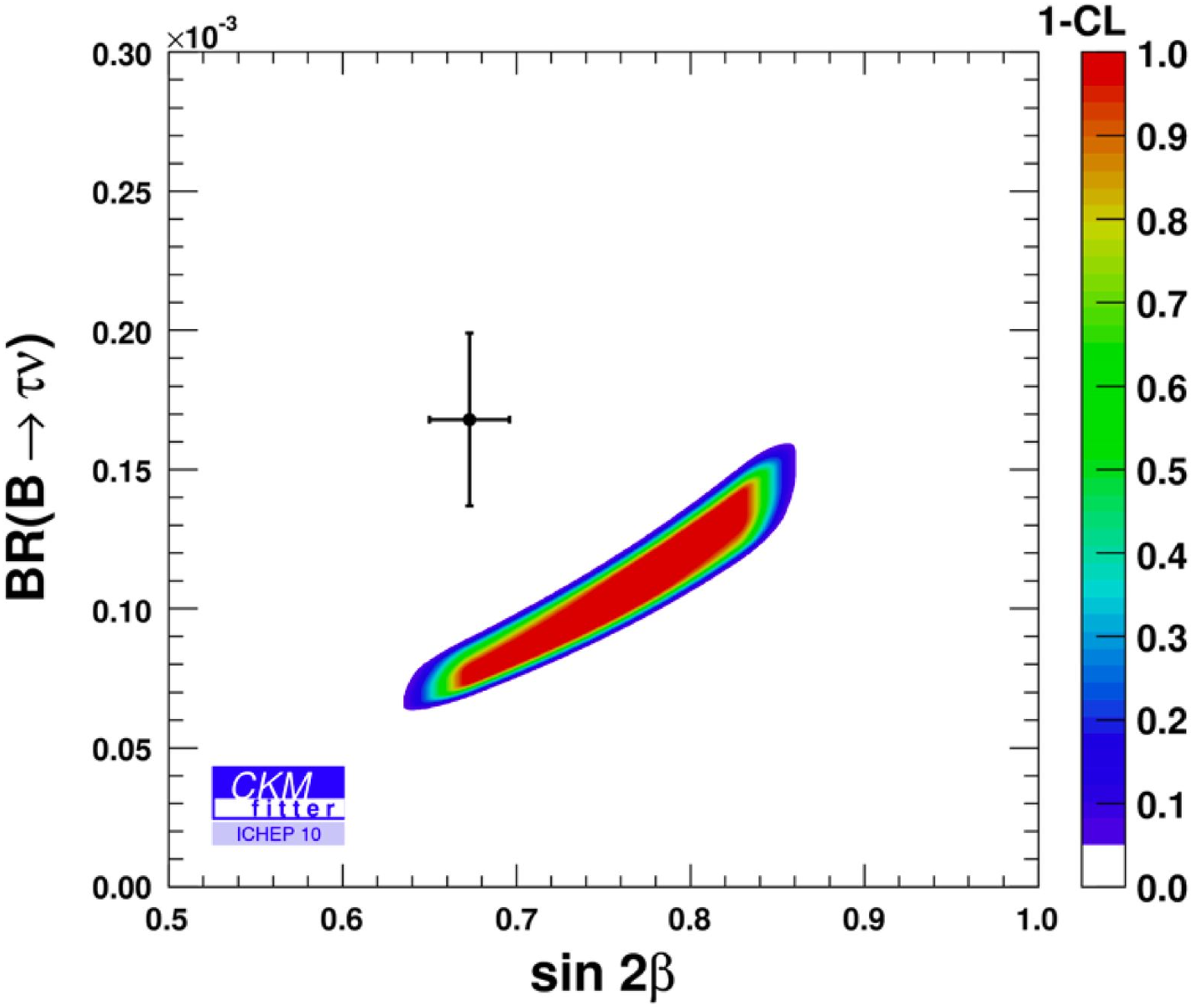}
\end{center}
\caption{Direct measurement versus indirect fit prediction for ${\cal B}(B \rightarrow \tau\nu)$ (left)
and for correlation (${\cal B}(B \rightarrow \tau\nu)$,$\sin\beta$) (right); cross corresponds to the 
experimental  values with $1 \sigma$ uncertainties~\cite{CKMfitter}}
\label{fig:CKMfitter12}
\end{figure}

It is worthwhile noting that there is some tension between the direct measurement and the indirect fit 
prediction  for ${\cal B}(B \rightarrow \tau\nu)$, see left plot of Fig.~\ref{fig:CKMfitter12}. The deviation is
$2.6 \sigma$. 
Moreover, as was pointed out by  the CKMfitter group~\cite{CKMfitter}, there is a specific correlation 
between $\sin \beta$ and \mbox{${\cal B}(B \rightarrow \tau\nu)$}  which is also a bit at odds, see right plot of Fig.~\ref{fig:CKMfitter12}.

%%%%%%%%%%%
%%%%%%%%%%

\begin{figure}[htb]
\begin{center}
\includegraphics[width=.53\textwidth]{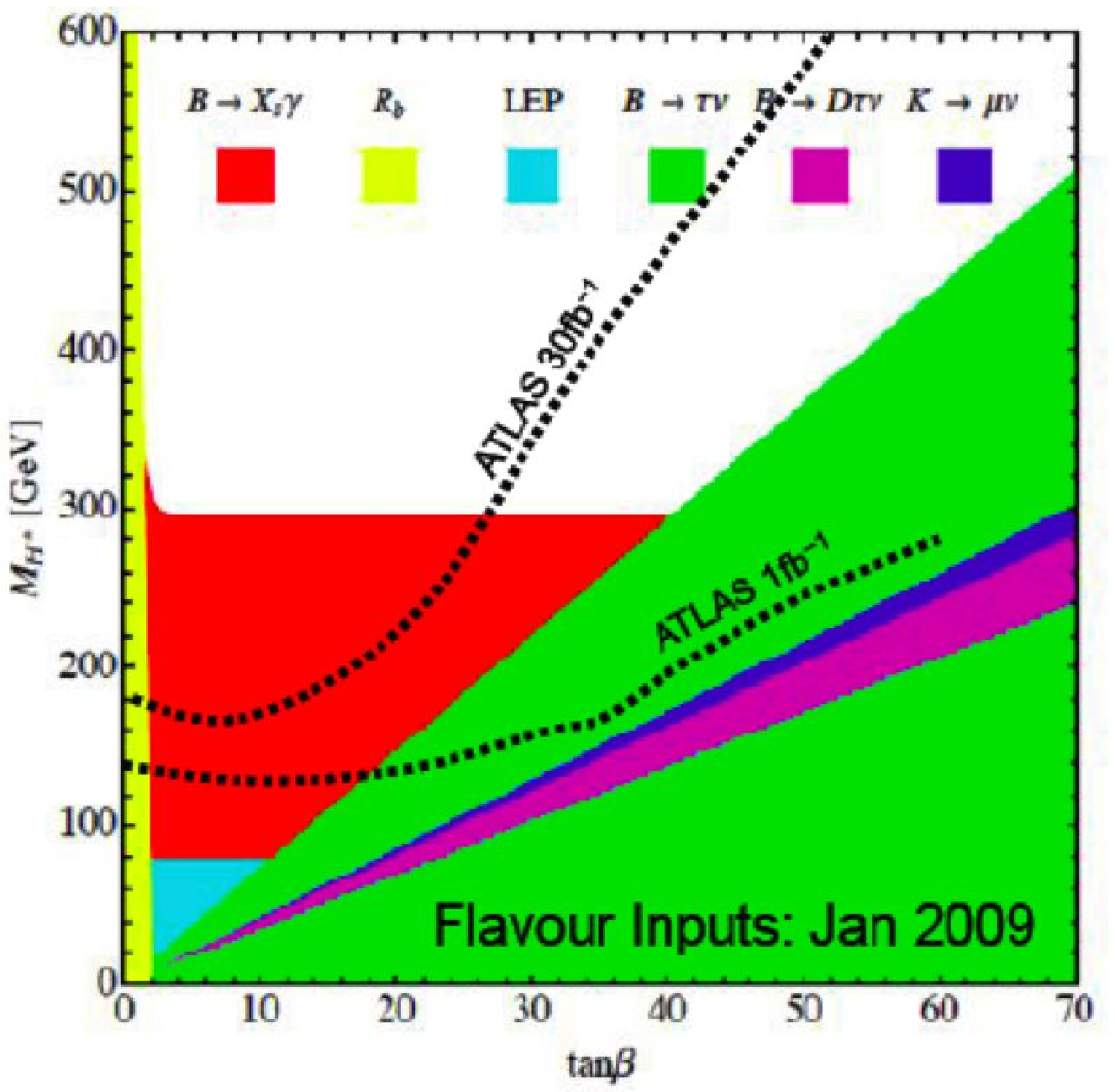}\includegraphics[width=.53\textwidth]{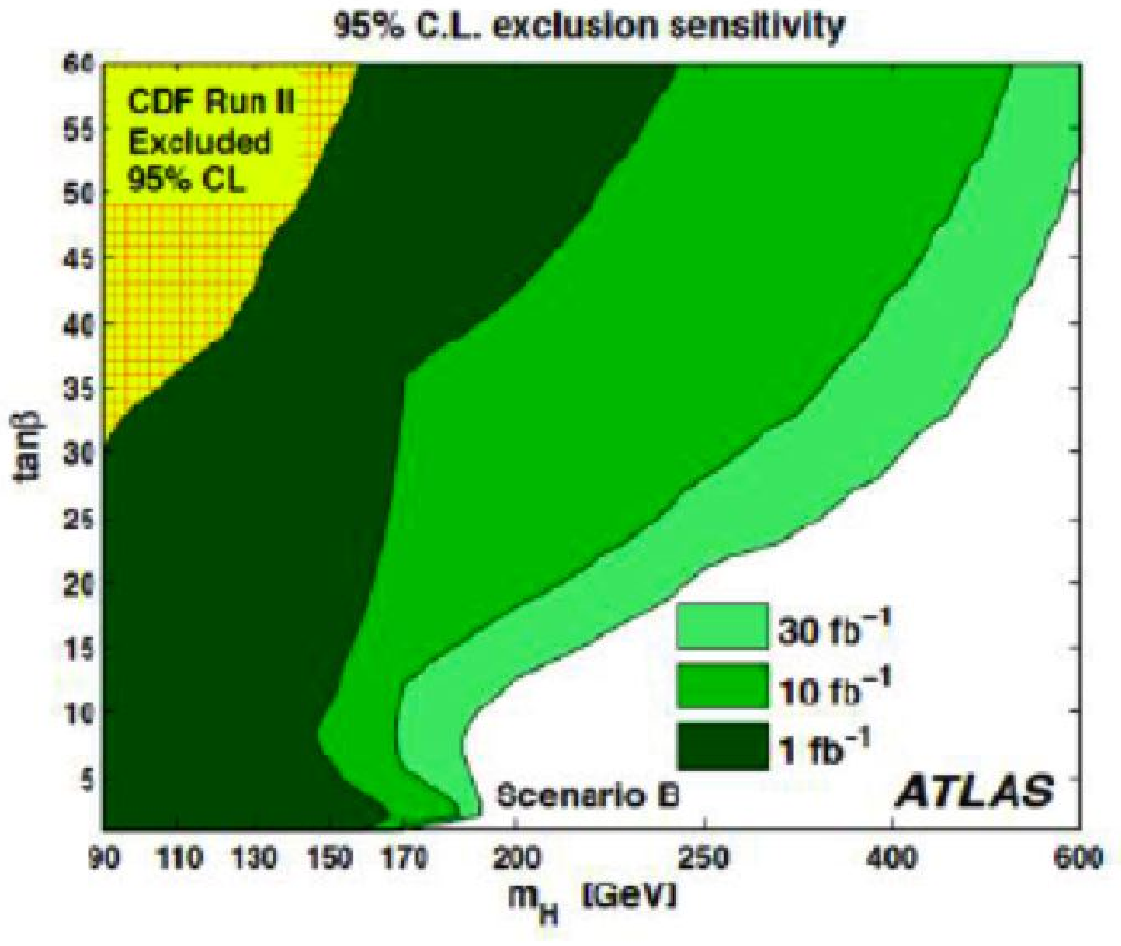}
\end{center}
\caption{Combined bound from all flavour observables with combined Higgs search constraint from ATLAS (left) and the original $95\%$ CL exclusion limits from  ATLAS~\cite{Bevan}}
\label{fig:charged2}
\end{figure}

A more recent combined analysis of all available bounds within the the THDM-II was presented 
in Ref.~\cite{Haisch:2008ar} and is shown in the left plot of Fig.~\ref{fig:charged2}: For $\tan \beta < 40$ 
the bound due to the $\bar B \rightarrow X_s \gamma$ decay is dominant, while for larger values
the tree-level process $B \rightarrow \tau \nu$  leads to the strongest bound. The latter  is less  model-dependent  and essentially survives also within the MSSM in contrast to the $\bar B \rightarrow X_s \gamma$ bound. Further analyses within different types of  THDMs can be found in 
Refs.~\cite{Mahmoudi:2009zx, Deschamps:2009rh}.

\begin{figure}[htb]
\begin{center}
\includegraphics[width=.59\textwidth]{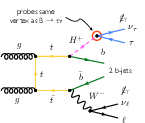}
\end{center}
\caption{ LHC observable tests out the same vertex as the decay $B \rightarrow \tau\nu$~\cite{Haisch}.}
\label{fig:charged1}
\end{figure}

Finally, the indirect NP reach via flavour data and the direct NP reach via the ATLAS and CMS experiments should  be compared: The expected $95\%$ CL exclusion limits of the LHC from 
the processes  
$gg/gb  \rightarrow t(b)  H^+, H^+ \rightarrow \tau \nu / t b$
~\cite{Aad:2009wy,Ball:2007zza}  (see also  Fig.~\ref{fig:charged1})
are shown in the right, but also in the left plot of Fig.~\ref{fig:charged2}; one finds that the present flavour constraints on the THDM-II are comparable and, therefore, nicely complementary to the 
expected exclusion limits  of the LHC. One needs around $10 {\rm fb}^{-1}$ at the LHC in order 
to reach {\it new} territory  not ruled out by the present flavour data.

%%%%%%%%%%%%%
%%%%%%%%%%%%%

%%%%%%%%%%%%%%%
%%%%%%%%%%%%%%%%%

\section{Minimal flavour violation hypothesis and CP issues}

The minimal flavour violation (MFV) hypothesis is a formal solution to the 
NP flavour problem. It assumes that the flavour and the CP 
symmetry are  broken as in the SM.  Thus, it 
requires that all flavour- and CP-violating interactions be 
linked to the known structure of Yukawa couplings 
(called $Y_U$ and $Y_D$ in the following).  
A renormalization-group-invariant definition of MFV 
based on a symmetry principle 
is given in~\cite{Chivukula:1987py,Hall:1990ac,D'Ambrosio:2002ex}; 
this  is mandatory for  a consistent  
effective field theoretical analysis
of NP  effects. 

In fact, a low-energy effective theory with all SM fields including one or two Higgs doublets can be constructed;
as the only source of $U(3)^5$ flavour symmetry breaking, 
the ordinary Yukawa couplings are introduced as background values of fields  
transforming  under the flavour group \mbox{(`spurions')}~\cite{D'Ambrosio:2002ex}. 
In the construction of the effective field theory,  operators with arbitrary 
powers of the dimensionless $Y_{U/D}$ have in principle to be  considered. 
However, the specific  structure of the SM,  with  its hierarchy 
of CKM matrix  elements and quark masses,  drastically reduces 
the number  of numerically  relevant operators. For example, it can be shown  
that in MFV models with one Higgs doublet, all FCNC  processes with  external 
$d$-type quarks are governed by 
the following combination of spurions due to the dominance of the top 
Yukawa coupling $y_t$:
\begin{equation}
(Y_U Y_U^\dagger)_{ij} \approx y_t^2  V^*_{3i} V_{3j}\,,  
\end{equation} 
where  a basis is used in which the  $d$-type quark Yukawa is diagonal. 

There are two strict predictions in this general 
class of models  which have to be  tested. First, the MFV hypothesis implies 
the   usual CKM relations between $b \to s$, $b \to d$, 
and  $s \to d$ transitions. For example, this relation allows 
for upper bounds  on new-physics effects in 
${\cal B}(\bar B \to X_d\gamma)$, and ${\cal B}(\bar B \to X_s \nu\bar \nu)$ using experimental data or bounds from ${\cal B}(\bar B  \to X_s\gamma)$, and 
${\cal B}(K \to \pi^+ \nu\bar \nu)$,  respectively. 
This emphasizes the need for 
high-precision measurements of $b \to  s/d$ , but also of 
$s \to  d$ transitions such as  the rare kaon decay   $K \to \pi \nu\bar\nu$. 
A systematic analysis of MFV bounds and relations for $\Delta F  =1$ transitions is given 
in Ref.~\cite{Hurth:2008jc}, for $\Delta F=2$ in Ref.~\cite{Bona:2007vi}. 
The usefulness of MFV-bounds/relations is obvious; any 
measurement beyond those bounds \mbox{indicate} the existence of new 
flavour structures.

It is well known that scenarios including  two Higgs doublets with large 
$\tan \beta = O(m_t/m_b)$  allow for  the unification of top and bottom 
Yukawa couplings, as predicted in grand-unified models~\cite{Hall:1993gn},  and 
for sizeable new effects in helicity-suppressed decay modes~\cite{Hamzaoui:1998nu,Babu:1999hn}.
There are more general MFV relations existing  in this scenario due 
to the dominant role of scalar operators. However, since  
$\tan \beta$ is large,  
there is a new combination of spurions numerically relevant in 
the construction of higher-order MFV effective \mbox{operators,} namely
\begin{equation}
(Y_D Y_D^\dagger)_{ij} \approx y_d^2  \delta_{ij}\,,
\end{equation} 
which invalidates the general MFV 
relation between $b \to s/d$ and $s \to d$ transitions.

Within the MFV hypothesis   the CKM phase is often assumed to be the only  source of CP violation. This implies that any phase 
measurement is  not sensitive 
to  new physics. But  flavour and CP violation can be treated separately. 
In fact, allowing for flavour-blind phases  there is   a
RG-invariant extension of the MFV concept possible, as was first discussed  in a   phenomenological analysis on CP-violating observables~\cite{Hurth:2003dk}\footnote{The MFV 
hypothesis with flavourblind phases  is sometimes called  $\overline{\rm MFV}$~\cite{Buras:2010zm}.}.  But in general  these phases  lead to non-trivial CP  effects,   which get however  strongly constrained by flavour-diagonal observables 
such as    electric dipole moments.

Nevertheless, more recently 
Batell and  Pospelov have given  a deeper insight  into the concrete EDM constraints on CP 
phases~\cite{Batell:2010qw}. 
They have shown that the large flavourblind CP phases which are compatible with the present EDM constraints almost exclusively contribute to the $B_s$ mixing.
In view of the present (slightly anomalous) data on $B_s$ mixing from the Tevatron 
experiments ~\cite{TevatronB1,TevatronB2} this is a very interesting new result~\cite{Lenz:2010gu}.

\section{Minimal flavour-violating THDM and stability issues}

In Ref.~\cite{Pich:2009sp,Tuzon:2010vt}, the authors propose the so-called {\it aligned} THDM
by fixing all the flavour matrices $X_i$ in Eq.~\ref{Lagrangian1} to be proportional to the 
corresponding Yukawa couplings:
\begin{equation}
X_{d1}= const_{d1} Y_D,\, \,\, X_{d2}= const_{d2} Y_D,\,\, \,X_{u1}= const_{u1} Y_U,\,\,\, X_{u2}= const_{u2} Y_U,
\label{MFVfirstorder}
\end{equation}
with real or flavourblind prefactors $const_i$. 
Comparing Eqs.~\ref{Massmatrices} and \ref{Couplings} in the aligned model, one finds 
that there are no FCNC at the tree level. 

But the aligned THDM is just  the most general minimal flavour-violating (MFV) \mbox{renormalizable} 
THDM, but  expanded to lowest order in the Yukawa couplings $Y_i$.   Following Ref.~\cite{D'Ambrosio:2002ex}, the most general MFV ansatz is  given by the expansion in the two left-handed spurions 
$Y_U Y_U^\dagger$ and  $Y_D Y_D^\dagger$  which were discussed in the last chapter: 
\begin{eqnarray}
X_{d1} &=& Y_D, \nonumber\\
X_{d2} &=& \epsilon_{0} Y_D + 
\epsilon_{1} Y_D  Y_D^\dagger Y_D                    
+  \epsilon_{2}  Y_U Y_U^\dagger Y_D + \ldots, \nonumber\\
X_{u1} &=& \epsilon^{'}_{0} Y_U + 
\epsilon^{'}_{1}  Y_U Y_U^\dagger Y_U  +  \epsilon^{'}_{2}  Y_D Y_D^\dagger Y_U + \ldots~, \nonumber\\
X_{u2} &=& Y_U.  \label{MFVall} 
\end{eqnarray}
The simple form of $X_{d1}$ and $X_{u2}$ can be assumed
without loss of generality by redefining the two spurions $Y_U$ and $Y_D$.
But if the higher-order terms in $X_{d2}$  and $X_{u1}$ are not included on the tree-level, as in 
ansatz~\ref{MFVfirstorder}, they are  automatically  generated by radiative corrections. This is assured by the RG invariance of the MFV 
hypothesis which is implemented by the {\it flavour}  $SU(3)^3$ symmetry. 
  Thus, the functional form in Eq.~\ref{MFVall} is preserved, only the coefficients 
$\epsilon_i$ and $\epsilon^{'}_i$ change and are related via the RG equations.  
In view of this, it is also clear that setting all coefficients to zero leads to heavy fine-tuning. 
Thus, there is no Yukawa alignment in general within the MFV framework.

In Ref.~\cite{Buras:2010mh}, the stability of the various tree-level implementations by flavourful and
flavourblind symmetries regarding flavour protection is discussed. 
In the MFV case, the FCNC induced by higher-order terms in the spurions 
are under control. Even when the coefficients in Eq.~\ref{MFVall} are of order $O(1)$ the 
expansion in the spurions is rapidly convergent due to small CKM matrix elements and 
small quark masses as was already shown in Ref.~\cite{D'Ambrosio:2002ex}.

This is not in general true in the case of  the implementation via exact flavourblind symmetries  if  there are  additional degrees of freedom at higher scales. Integrating out  the latter,  
one can easily  construct higher-dimensional operators which are $Z_2$ invariant, but   which destroy the flavour protection:
\begin{eqnarray}
 L_Y^{\rm d > 4} &=&  \frac{c_1}{\Lambda^2} \bar Q_L X^{(6)}_{u1} U_R H_2 |H_1|^2
+ \frac{c_2}{\Lambda^2} \bar Q_L X^{(6)}_{u2} U_R H_2 |H_2|^2 \nonumber  \\
&& + \frac{c_3}{\Lambda^2} \bar Q_L X^{(6)}_{d1} D_R H_1 |H_1|^2
+ \frac{c_4}{\Lambda^2} \bar Q_L X^{(6)}_{d2} D_R H_1 |H_2|^2~,
\end{eqnarray}
These operators are $Z_2$ exact in the sense of the Type-II model ($H_1 \leftrightarrow H_1$ and $D_R \leftrightarrow -D_R$), but after electroweak symmetry breaking they  induce new 
FCNC.  With $c_i = O(1)$ and the new physics scale $\Lambda = O(1 {\rm TeV})$ one finds too large FCNC inconsistent with present flavour data~\cite{Buras:2010mh}.
Further protection via the MFV hypothesis is needed. This problem already occurs  
in the case of  one Higgs doublet~\cite{Giudice:2008uua,Agashe:2009di,Azatov:2009na}.

A similar argument for the implementation of the tree-level  condition using the Peccei-Quinn $U(1)$ is 
valid.  However, in contrast to the $Z_2$ symmetry, the Peccei-Quinn symmetry must be explicitly broken in other sectors of the theory to avoid a massless pseudoscalar Higgs field. The {\it spontaneous}
breaking via the vev of $H_2$ would imply a Goldstone boson. In general, the explicit breaking terms induce too large FCNC~\cite{Buras:2010mh}.

\section{Future opportunities}

There are  great experimental opportunities in flavour physics in the near future.   
LHCb~\cite{LHCb} has finally started taking data and promises to overwhelm many
$B$ factory results. 
In addition, two Super-$B$ factories, Belle II at KEK~\cite{Aushev:2010bq, Abe:2010sj}
and 
SuperB in Italy~\cite{O'Leary:2010af, Hitlin:2008gf, Bona:2007qt},  have been approved  and partially funded to accumulate two 
orders of magnitude larger data samples. 
The Super-$B$ factories are  Super-{\it Flavour} factories:  Besides precise $B$ measurements 
they allow for precise analyses of CP violation in charm and of lepton flavour-violating modes 
like $\tau \rightarrow \mu\gamma$ (for more details see Ref.~\cite{Browder:2007gg}).

%%%%%%%%%%
\begin{figure}[t]
  \begin{center}
    \includegraphics[width=0.31\textwidth]{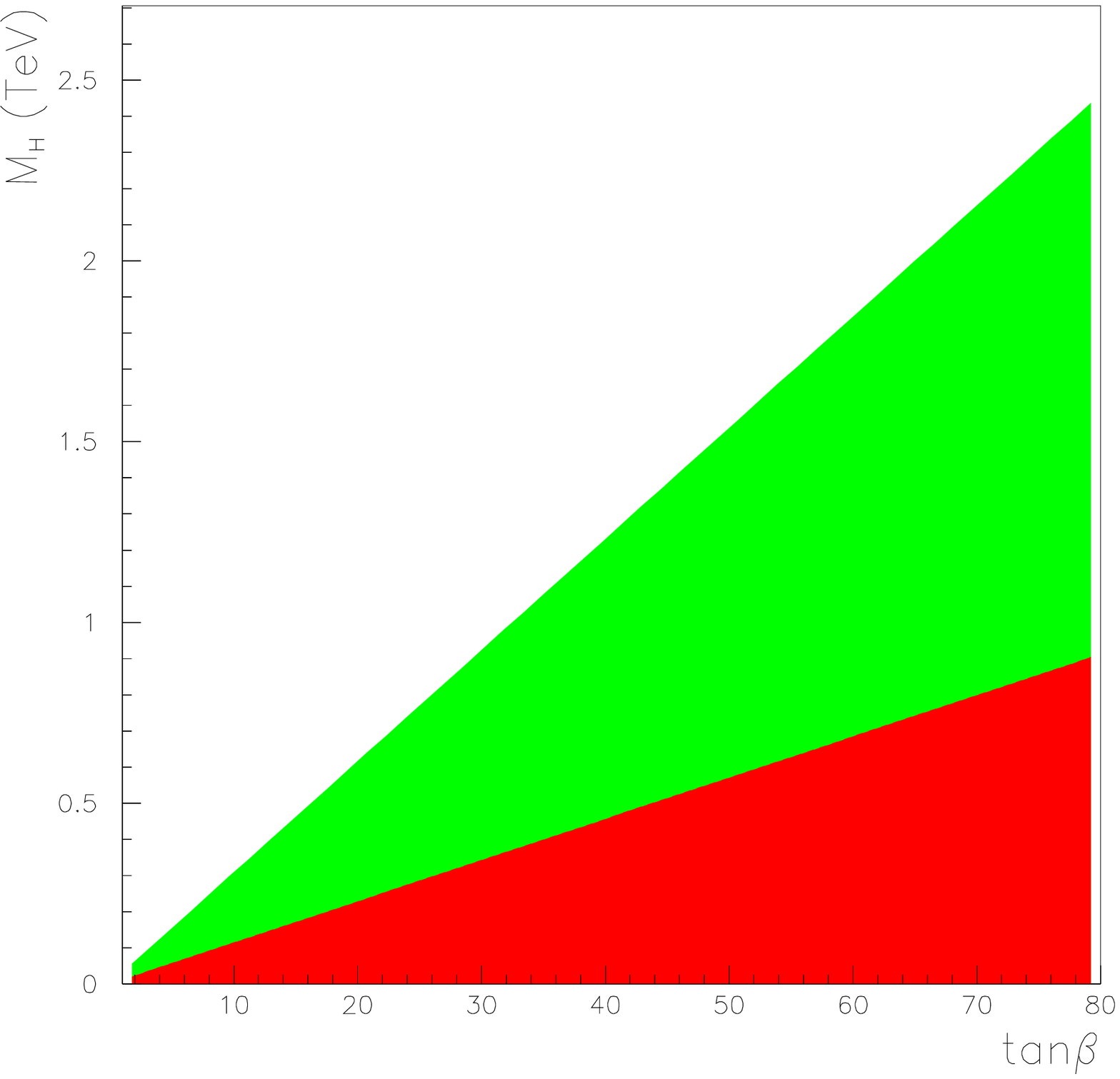} 
    \includegraphics[width=0.31\textwidth]{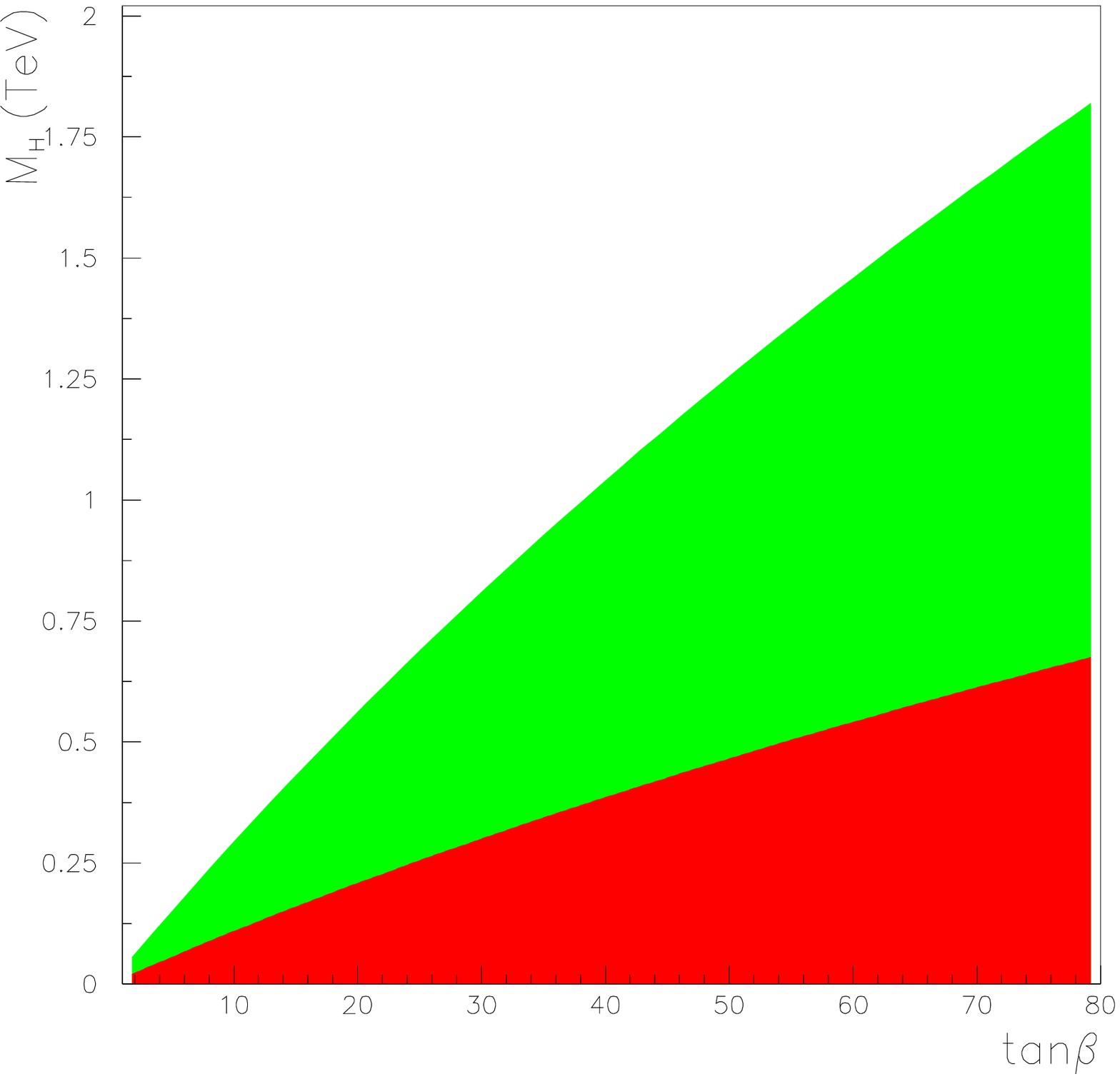} 
    \caption{
      Exclusion regions at 95\% probability 
      in the $M_{H^\pm}$--$\tan \beta$ plane for the THDM-II (left) 
      and the MSSM (right) obtained
      assuming the SM value of ${\cal B}(B \rightarrow \tau\nu)$ measured 
      with $2 \ {\rm ab}^{-1}$ (dark (red) area,  $B$ factories)  and
      $75 \ {\rm ab}^{-1}$ (dark (red) + light (green) area, Super $B$ factories)~\cite{Browder:2007gg}. 
     }
    \label{fig:btaunu}
  \end{center}
\end{figure}
%%%%%%%%%%%

Regarding the measurement of clean $B$ modes, 
the Super-$B$ factories will push the experimental precision to its limit. For example,
the present  experimental error of  ${\cal  B}(B \rightarrow \tau\nu)$ discussed in Sect.~3  will be reduced from $20\%$
down to $4\%$. Thus, the NP reach of this observable will  siginificantly 
improve; exclusion regions within the THDM-II and the MSSM are shown in Fig.~\ref{fig:btaunu}. \\

%%%%%%%%%%%%%%%%%%%%%%%%%%%%%%%%%%%%%%%%%%%%%%%%%%%%%%%%%%%%%%%%%%%%%%%%%%%

\section*{Acknowledgement} TH thanks the organizers of the workshop for the interesting and valuable meeting,  Leonardo Vernazza for a careful reading of the manuscript, and the CERN theory group for its  hospitality during his regular visits to CERN where
part of this work  was written \\

\newpage

%\begin{multicols}{2}

\end{document}